\documentstyle[12pt,prl,aps]{revtex}

\begin{document}
\draft
\preprint{}
\title{Super Virasoro Ghost Algebra using the Conformal Ghosts}
\author{B. B. Deo\footnote{bbdeo@yahoo.com}}
\address{Physics Department, Utkal University, Bhubaneswar 751 004}
\maketitle{}
\vskip 0.5in
\begin{abstract}
Superconformal ghost current generators of conformal dimension $3/2$ are constructed using the conformal ghosts and anticommuting infinite dimensional gamma matrices of the Clifford algebra. The super-Virasoro algebra for the ghosts in both the N.S. and R sectors are presented. The anomaly terms in both cases are deduced using Jacobi identity.
\end{abstract}
\vskip 0.5in
The commutator of bosonic string Virasoro operators \cite{virasoro},
which are Fourier modes of energy momentum tensor, contain quantum
mechanical anomalous terms. In covariant formulation, these terms are
cancelled by introducing the conformal ghosts \cite{green}. The ghost
quanta are labelled by $c_n$ and the antighosts by $b_n$. `$n$' is an
integer. The canonical anticommutation relations satisfied by them, are
\begin{equation}
\{ c_m , b_n \} = \delta_{m+n} 
\end{equation}
\begin{equation}
\{ c_m , c_n \} = \{ b_m , b_n \} = 0
\end{equation}
The Virasoro ghost operator is
\begin{equation}
L_m^{gh} = \sum_{ n = - \infty}^{\infty} (m - n) b_{m+n} c_{-n}
\end{equation}
The antighost $b$ has a conformal dimension $+2$ where as the ghost
$c$ has conformal dimension $-1$, so that
\begin{equation}
\left [ L_m^{gh} , b_n \right ] = (m-n) b_{m+n}
\end{equation}
\begin{equation}
\left [ L_m^{gh} , c_n \right ] = -(2m+n) c_{m+n}
\end{equation}
Virasoro algebra  is \cite{green}
\begin{equation}
\left [ L_m^{gh} , L_n^{gh} \right ] = (m-n) L_{m+n}^{gh} + A^{gh}(m) \delta_{m+n}
\end{equation}
The last terms is the ghost anomaly term which cancels the normal
anomaly $\frac{\bf c}{12} (m^3 - m)$ when the central charge ${\bf c}=26$ and $a=1$ \cite{green} i.e. $A^{gh}(m) = - \frac{26}{12}(m^3 - m) = - A^{gh} (-m)$.

We shall have a super Virasoro algebra with ${\bf c}=26$ if we can
construct the ghost current generators $F_r^{gh}$ or $G_r^{gh}$ in
Ramond \cite{ramond} (where $r$ is integral)or Beveu-Schwarz \cite{schwarz} (where $r$ is half integral) respectively. In Ramond sector we should have
\begin{equation}
\left [ L_m^{gh} , F_r^{gh} \right ] = \left ( \frac{m}{2} - r \right ) F_{m+r}^{gh}
\end{equation}
\begin{equation}
\{ F_r^{gh} , F_s^{gh} \} = 2 L_{r+s}^{gh} + B^{gh} (r) \delta_{r+s}
\end{equation}
$B^{gh}(r)$ is the ghost current anomaly term.

The current generator must have a conformal dimension $3/2$. To
construct such an object we use the gamma matrices of the  Clifford \cite{clifford} algebra.

Consider a representation of Clifford algebra in infinite dimensions
\begin{equation}
\gamma_a \gamma_b + \gamma_b\gamma_a = \{ \gamma_a, \gamma_b \} = 2 \delta_{ab} {\bf I}, \; \; a,b = 1,2 \cdots \infty
\end{equation}
Multiplying these matrices we obtain the matrix `$\gamma_n$' of degree $n$,
\begin{equation}
\gamma_n = \gamma_{a1} \gamma_{a2} \cdots \gamma_{an} , \; \; a1 < a2 \cdots < an
\end{equation}
Since $\gamma_a^2 = 1$, we have
\begin{equation}
\gamma_n^2 = (-1)^{1/2 n(n-1)} {\bf I}
\end{equation}
The sign factor arises because of the reordering the $\gamma$'s to
become equal to the neighbour.  When $n$ is odd, we can choose a set
of gamma matrices which satisfy
\begin{equation}
\{ \gamma_{2j-1} , \gamma_{2 j^\prime - 1} \} = 2(-1)^j \delta_{j j^\prime} {\bf I}
\end{equation}
We construct the sum of such product matrices,
\begin{equation}
\Gamma_n = \sum_{j=1}^{n} c_{2j-1} \gamma_{2j - 1}
\end{equation}
\begin{equation}
\{ \Gamma_n , \Gamma_m \} = \sum_{j=1}^n c_{2j-1}^2 (-1)^j 2 \delta_{m,n} {\bf I}
\end{equation}
In applying these results to the problem at hand,  we shall need the values $< \alpha \mid \Gamma_{|n|}^{\alpha\beta} \mid \beta >$ where $\alpha, \beta$ are states normalised such that $< \alpha \mid \alpha > = 1$ and $\mid \alpha > < \beta \mid = \delta_{\alpha \beta}$. For simplicity in notation, we have also used the $\Gamma_n$, to indicate
\begin{equation}
\Gamma_n = < \alpha \mid \Gamma_{|n|}^{\alpha\beta} \mid \beta >
\end{equation}
such that
\begin{equation}
\{ \Gamma_n , \Gamma_m \} = 2 \sum_{j=1}^{|n|} c_{2j-1}^2 (-1)^j \delta_{m,n}
\end{equation}
In particular
\begin{equation}
\Gamma_n^2 = \sum_{j=1}^{|n|} (-1)^j c_{2j-1}^2
\end{equation}
Since the coefficients $c_{2j-1}^2$ are at our choice, we choose in such a fashion that every
\begin{equation}
(-1)^j c_{2j-1}^{2} = - \epsilon(n)
\end{equation}
This leads to 
\begin{equation}
\Gamma_n^2 = -n
\end{equation}
Since $\Gamma_n$ is now proportional to $\sqrt{-n}$, the ghost current
having conformal dimension $3/2$, can  be guessed.
\begin{equation}
F_r^{gh} = \sum_{n= - \infty}^\infty \Gamma_n b_{n+r} c_{-n}
\end{equation}
Using the commutation relations (4) and (5) and the anticommutation
relation (16), we deduce that
\begin{eqnarray}
[ L_m^{gh} , F_r^{gh}] \ = && ( \frac{m}{2} - r) \sum_n \Gamma_n b_{n+m+r} 
c_{-n} \nonumber \\
 && + \sum_n \left ( (\frac{m}{2} -n) \Gamma_n - \Gamma_{n\oplus m}
   (m-n) \right ) b_{n+m+r} c_{-n}
\end{eqnarray}
and
\begin{equation}
\{ F_r^{gh}, F_s^{gh} \} = \sum_n ( \Gamma_{n\oplus r} \cdot \Gamma_m 
+ \Gamma_{n\oplus s} \cdot \Gamma_m ) b_{n+r+s} c_{-n}
\end{equation}

$\Gamma_{n\oplus m}$ will be the result of shifting the $n$ in
$\Gamma_n$ by an amount $m$ keeping the gamma within the calss
$\Gamma_n$ defined by equations (12), (13) and (14). The product, like the ones
occuring above,
\begin{equation}
\Gamma_{n\oplus m/2} \cdot \Gamma_n = \sum_{j=1}^{|n\oplus m/2|}
\sum_{j^\prime = 1}^{|n|} c_{2j - 1} c_{2j^\prime - 1}
\gamma_{2j -1} \gamma_{2j^\prime  - 1}
\end{equation}
\begin{equation}
= \sum_{j=1}^{|n\oplus m/2|} \sum_{j^\prime = 1}^{|n|} c_{2j-1}
c_{2j^\prime - 1} \left ( \frac{1}{2} \{ \gamma_{2j-1} ,
  \gamma_{2j^\prime - 1} \} + \frac{1}{2} [ \gamma_{2j - 1},
  \gamma_{2j^\prime - 1} ] \right )
\end{equation}
It will eventually turn out that we need consider $j_{\rm max} =
j^\prime_{\rm max}$. So the second, the commutator term, vanishes as
can be seen by interchanging $j \leftrightarrow j^\prime$.

With the help of the Heavyside theta function, the first term becomes
\begin{equation}
\sum_j \theta (|n| -j) \theta (|n\oplus m/2| -j) (-1)^j c_{2j-1}^2
\end{equation}
If $|n\oplus m/2|$ is greater than $|n|$, then it is redudant; and 
the shift produces no change. This is not true.  For the shift to be meaningful $|n\oplus 
m/2|$ should be less than $|n|$. The first theta function is
redudant. The sum becomes
\begin{equation}
\sum_{j=1}^{|n\oplus m/2|} (-1)^j c_{2j-1}^2
\end{equation}
with $|n\oplus m/2| < n$. Let us look at the last terminating term
with the shift involving the gamma matrices namely
\begin{equation}
\gamma_{2(n\oplus m/2) -1} = \gamma_{(2n-1)\oplus m}
\end{equation}
If the term is to going to contain less than the product of $n$
$\gamma$-matrices, $m$ pairs of $\gamma$-matrices must be becoming
unity i.e. the additional $m$ gamma matrices of $\gamma_{(2n-1)\oplus
  m}$ must be equal to $\gamma_{2n-2m-1}$ multiplied by successively
equal neighbouring $m$ pairs of gamma matrices. 

So we are led to define the end term to be
\begin{equation}
\gamma_{(2n - 1)\oplus m} = \gamma_{2n-2m-1}
\end{equation}
and $j$ in equation (26) has the maximum value $|n-m|$. As a result,
the shifted $\Gamma_{n\oplus m/2}$ satisfies the relation
\begin{equation}
\Gamma_{n\oplus m/2} \cdot \Gamma_n = m - n
\end{equation}
and it follows that (by shifting $n$ by $m/2$),
\begin{equation}
\Gamma_{n\oplus m} \Gamma_{n\oplus m/2} = m - (n+m/2) = m/2 - n
\end{equation}
and
\begin{equation}
\Gamma_{n\oplus m} \Gamma_n = 2m-n
\end{equation}
Using (29) and (30) we see that the last two unwanted terms in
equation (21) cancel and 
\begin{equation}
\left [ L_m^{gh} , F_r^{gh} \right ] = \left ( \frac{m}{2} - r \right
) F_{m+r}^{gh}
\end{equation}
The anticommutator, equation (22) is now
\begin{equation}
\{ F_r^{gh}, F_s^{gh} \} = 2 \sum_n (r+s-n) b_{n+r+s} c_{-n} = 2
L_{r+s}^{gh} + B^{gh}(r) \delta_{r+s}
\end{equation}

Let us now calculate the anomaly term for euqation (33). $F_o^{gh}$
has no normal ordering ambiguity as $\Gamma_o = 0$. So $F_o^{gh^2} =
L_0^{gh}$. Further $F_r^{gh} = \frac{2}{r} [ L_r^{gh}, F_o^{gh}]$. 
Using Jacobi indentity,
\begin{equation}
\left [ \{F_r, F_s\}, L_m\right] + \{ [ L_m, F_r],F_s \} + \{ [ L_m,
F_s],F_r\} = 0
\end{equation}
we get
\begin{equation}
\{ F_r^{gh}, F_{-r}^{gh} \} = \frac{2}{r} \{ [ L_r, F_0], F_{-r} \}
= 2 L_o^{gh} + \frac{4}{r} A^{gh}(r) = 2 L_o^{gh} + B^{gh}(r)
\end{equation}
As a result
\begin{equation}
B^{gh} (r) = - \frac{\bf c}{3} (r^2 -1), \; \; r \neq 0
\end{equation}
There is difficulty in going over to the NS sector, as the $r$ in the 
$G_r^{gh}$ takes half integral values whereas the conformal ghosts are 
defined for integers. To circumvate this difficulty we introduce an
integer $r + 1/2t$ where $t$ is an integer and define
\begin{equation}
G_r^{gh} = Lt_{t \rightarrow 0} \sum_n \Gamma_n b_{n+r+t/2}
c_{-n}
\end{equation}
with the understanding that the limit $t\rightarrow 0$ will be taken
after all calculations are over. We  reproduce the whole
Super Virasoro algebra expect for the anomaly term like equation (34)
and (35). Here, again, the Jacobi Identity equation (33) for the $G$'s can 
be used to show that
\begin{equation}
B^{gh}(r) = \frac{A^{gh}(2r)}{2r} = - \frac{\bf c}{3} (r^2 - \frac{1}{4})
\end{equation}
Collecting the results, we have the super Virasoro ghost algebra
\begin{equation}
[ L_m^{gh} , L_n^{gh} ] = (m-n) L_{m+n}^{gh} - \frac{\bf c}{12} (m^3-m)
\delta_{m+n}
\end{equation}
N.S.:
\begin{equation}
[L_m^{gh}, G_r^{gh}] = \left( \frac{m}{2}-r\right ) G_{m+r}^{gh}
\end{equation}
\begin{equation}
\{G_r^{gh}, G_s^{gh} \} = 2 L_{r+s} - \frac{\bf c}{3} (r^2 - 1/4)
\delta_{r+s}
\end{equation}
R:
\begin{equation}
[L_m^{gh}, F_r^{gh}] = ( \frac{m}{2} - r) F_{m+r}^{gh}
\end{equation}
\begin{equation}
[ F_r^{gh}, F_s^{gh} ] = 2 L_{r+s} - \frac{\bf c}{3} (r^2 - 1)
\delta_{r+s} , \; r \neq 0
\end{equation}
The super conformal ghosts have never been used. We hope that this will
stimulate research to findq supersymmetric string theories with central
charge 26 in particular \cite{deo}.

\end{document}